\documentclass[showkeys,showpacs,prd]{revtex4}
\usepackage{graphicx}
\usepackage{amsmath}
\usepackage[cp1251]{inputenc}
\usepackage[russian]{babel}
\usepackage{amssymb}
\usepackage{color}

\begin{document}

\title{Phase transition for gluon field: a qualitative analysis}
\author{Vladimir Dzhunushaliev}
\email{v.dzhunushaliev@gmail.com}
\affiliation{
Dept. Theor. and Nucl. Phys., KazNU, Almaty, 010008, Kazakhstan}

\begin{abstract}
The phase transition for SU(3) gauge field (without quarks) is considered. It is shown that the phase transition is due to the fact that at high temperatures the partition function should be calculated as for a gas of  gluons, whereas at low temperatures as the sum over energy levels of correlated quantum states of SU(3) gauge field. A correlated quantum state for strongly interacting fields is defined as a nonperturbative quantum state of strongly interacting fields. The energy spectrum of these quantum states are discrete one. A lower bound of the phase transition temperature by comparing of the average energy for the perturbative and nonperturbative regimes is estimated (for glueball being in thermal equilibrium with the thermostat). It is shown that this quantity is associated with a  mass gap. In a scalar model of glueball its energy is  calculated. It is shown that this energy is the mass gap. If we set the glueball mass $ \approx 1.5 \cdot 10^3$ Mev then it is found that the corresponding value of coupling constant lies in the nonperturbative region.
\end{abstract}


\maketitle

\section{Introduction}

One of the most interesting predictions of QCD is a phase transition at some critical temperature to a new phase of strongly interacting fields: the quark-gluon plasma. The quark-gluon plasma is a state of strongly interacting matter (with the quarks and gluons) where they are not longer confined to color neutral entities of hadronic size. The quark-gluon plasma  is a state of matter at temperatures above the phase transition temperature and this state is characterized by a weak coupling constant $q < 1$. At the temperatures below the phase transition temperature we have strong coupling constant $g \gtrsim 1$ and we should use nonperturbative technique to describe quantum fields in this region. The various methods can be employed to describe the thermodynamics of the high temperature quark-gluon plasma (for review, see \cite{Blaizot:2003tw} - \cite{Satz:2011wf} and references therein). 

Currently one of the biggest problems 
%(in our opinion it is a challenge) 
in quantum field theory is the problem of a nonperturbative quantization. The problem is that for strongly interacting fields  we cannot apply quantization recipes designed for weakly interacting fields. From a mathematical point of view it means that in this case the Feynman diagram technique cannot be applied. From a physical point of view it means that quantum strongly interacting fields cannot be presented as a cloud of interacting quanta. Rather such field can be compared with a turbulent fluid flow. In this flow there is a statistically fluctuating field of velocities. The velocities in the two sufficiently close points are correlated with each other. It was exactly what we have for strongly interacting quantum fields: the value of the quantum field in the two close enough spacelike points are correlated with each other. It means that the correlation function of these fields (2-point Green's function) in two sufficiently close spacelike points is nonzero. It has been noted by W. Heisenberg in Ref. \cite{heisenberg}. Evidently only one difference between quantum fields and turbulent fluid is that in the quantum case corresponding quantum states will be quantized. As it should be in quantum theory.

The quantum states describing the distribution of strongly interacting quantum fields will be called as \textcolor{blue}{\emph{quantum correlated states of strongly interacting fields}}. Briefly, \textcolor{blue}{\emph{quantum correlated states}}. For simplicity, in the future we will consider only the SU(3) non-Abelian gauge field without quarks.

The main goal of our research is a qualitative explanation of the phase transition for gluon field as a transition from the statistics of a gas of nonperturbative interacting gluons to the statistics nonperturbative correlated quantum states of strongly interacting SU(3) gauge field.

\section{Qualitative evaluation of the phase transition temperature}

Thus we assume that in the strong coupling regime, i.e. when the coupling constant $g \gtrsim 1$, the quantum SU(3) gauge field forms a kind of correlated distribution of the fluctuating field in space. Such distributions form a discrete spectrum. The discreteness  means that the energy spectrum of such distributions is discrete one. Note the following features of this point of view: the energy density of such distribution of  quantum field have to tend fast enough to zero at infinity that the energy of the field in whole space would be finite. It means that such fields are finite in space. For example, it may be a glueball. In this case above mentioned discrete spectrum of energy means that there is a mass spectrum of the glueball. The lowest energy will be called a mass gap.

If bring the glueball in thermal contact with the thermostat then the glueball can be considered as a statistical system described by a certain temperature and the fluctuating energy.

As it is known the problem of phase transition in the gluon plasma is that the behavior of the plasma for small and large temperatures is essentially different. General expectation is that this difference is due to the fact that at low temperatures the quantum description of the gluon field should be carried out by a non-perturbative way while at high temperatures the description of quantum gluon field should be carried out by the perturbative way.

For statistical calculations at high temperatures we will apply the statistics of gas of interacting gluons. In the limit $g \rightarrow 0$ we have the partition function 
\begin{equation}
	Z_{pert} =
	\sum \limits_{n=0}^\infty e^{ - \frac{\hbar \omega}{kT}}
\label{10}
\end{equation}
where $\hbar \omega$ is the energy of noninteracting gluons. After that by the standard way we obtain the Planck's formula for energy density (as for photons)
\begin{equation}
	u(\omega,T) = \dfrac{\hbar}{\pi^2 c^3}
	\dfrac{\omega^3 }{e^{\hbar \omega/kT} - 1}.
\label{20}
\end{equation}
At small temperatures the statistical sum looks as follows
\begin{equation}
	Z_{nonpert} =
	\sum \limits_{n=0}^\infty e^{ - \frac{E_n}{kT}}
\label{30}
\end{equation}
here $E_n$ is the energy of $n$-th \textcolor{blue}{\emph{quantum correlated state}}; $E_0 = \Delta$ is a mass gap. The calculation of $E_n$ is a problem of nonperturbative  quantization of strongly nonlinear fields (in our case it is SU(3) non-Abelian gauge field). This problem is extremely difficult. Below we will give an approximate calculation of the bound state of glueball using a scalar approximation for 2nd and 4th  Green's  functions.

Thus at low temperatures we should use the expression \eqref{30} but for high temperatures the expression \eqref{10}. The phenomenon of phase transition is that at some temperature the mean values of energy corresponding to statistical sums \eqref{10}  and \eqref{30} are of the same order of magnitude. It means that there is a transition from the description of a quantum field on nonperturbative language to the description of a quantum field on perturbative language. It is essentially important to note that \textcolor{blue}{\emph{energy of $E_n$ cannot be calculated using Feynman diagram technique}}.

Let us consider glueball being in thermal equilibrium with the thermostat. To describe the phenomenon of phase transition in this thermodynamic system we will consider two limiting cases. In the 1st case we have the glueball filled with weakly interacting gluons (perturbative case). The situation in this case is completely similar to statistics of the photons filling a certain box. With one exception: photons do not interact with each other. In the 2nd case the glueball can be in one of $E_n$ energy levels. The probability to be in one of $E_n$ is defined by standard methods of statistical mechanics.

The calculation of the mean value of energy density for weakly interacting gluon gas in the limit when the coupling constant $g \rightarrow 0$ is made in the usual way (i.e.  perturbative way) and gives us the standard formula for the total density of the Planck energy
\begin{equation}
	U = \sigma T^4 .
\label{40}
\end{equation}
The total energy of gluon gas in such glueball is 
\begin{equation}
	E_{pert} \sim \sigma T_c^4 r_0^3
\label{50}
\end{equation}
where $r_0$ is the characteristic size of the glueball and $T_c$ is the phase transition temperature. We assume that the characteristic size of the glueball coincides with the characteristic size of the nucleus $r_0 \approx 10^{-15}$m = 1fm.

The calculation of statistical mean value of energy for glueball filled with a strongly interacting gluon field gives us 
\begin{equation}
	\left\langle E \right\rangle_{nonpert} = 
	\dfrac{1}{Z_{nonpert}}
	\sum \limits_{n=0}^\infty E_n e^{ - \frac{E_n}{kT}}.
\label{60}
\end{equation}
The phase transition temperature separating perturbative physics from nonperturbative physics can be estimated as follows
\begin{equation}
	\left\langle E \right\rangle_{nonpert} \approx
	E_{pert}.
\label{70}
\end{equation}
Unfortunately we can not calculate 
$\left\langle E \right\rangle_{nonpert}$ but since the energy spectrum
\begin{equation}
	E_0 \leq E_1 \leq E_1 \leq \cdots
\label{80}
\end{equation}
we can take the lower bound of this temperature as follows
\begin{equation}
	\left\langle E \right\rangle_{nonpert} \geq
	E_0 = \Delta
\label{90}
\end{equation}
where $\Delta$ is the mass gap. Thus we obtain the lower bound
\begin{equation}
	\sigma T_c^4 r_0^3 \geq \Delta
\label{100}
\end{equation}

\section{Mass gap for glueball}

Now we want to show how it is possible to calculate approximately mass gap for glueball. Shortly the idea is as follows (in details it is possible to familiarize with calculations in Ref. \cite{Dzhunushaliev:2011we, Dzhunushaliev:2010ab}): In the scalar model of glueball the  Lagrangian of SU(3) non-Abelian gauge field is approximately represented in the form of  Lagrangian of two scalar fields $\phi, \chi$. One of these fields approximately describes the 2-point Green function of the gauge field $A^a_\mu \in SU(2) \subset SU(3), a=1,2,3$; and another scalar field, respectively, the 2-point Green function of the gauge field $A^m_\mu \in SU(3)/SU(2)$ where $SU(3)/SU(2)$ is a coset. 4-point Green's function is a bilinear combination of 2-point Green's functions. Similar methods are used for the description of turbulent fluid flow (see, for example,  \cite{ilyushin}). Thus these scalar fields describe the quantum fluctuations of the gluon field in the nonperturbative regime.

The Lagrangian for these fields is as follows 
\begin{equation}
	g^2 \mathcal L_{eff} = \frac{1}{2} \left( \partial_\mu \chi \right)^2 +
	\frac{1}{2} \left( \partial_\mu \phi \right)^2 -
	\frac{\lambda_1}{4} \left(
		\chi^2 - m_1^2
	\right)^2 -
	\frac{\lambda_2}{4} \left(
		\phi^2 - m_2^2
	\right)^2 - \frac{\lambda_3}{2} \phi^2 \chi^2 +
	\text{const}
\label{mg-10}
\end{equation}
where $g$ is the gauge coupling constant of SU(3) gauge field; $\lambda_{1,2}$ and  $m_{1,2}$ are constants. Corresponding field equations for the spherical symmetric case is given in Eq's \eqref{mg-40} \eqref{mg-50} where the constant $m_2$ is chosen in such a way that at infinity the energy density would tend to zero. These equations are considered as a nonlinear eigenvalue problem for eigenfunctions $\phi(r), \chi(r)$ and eigenvalue $m_2$. In Ref. \cite{Dzhunushaliev:2011we} a spherically symmetric solution describing the ball filled with quantum fluctuations of the gluon field is obtained. This solution is interpreted as a glueball.

Glueball energy is calculated as
\begin{equation}
	E_{gl} = \dfrac{4 \pi}{g^2} \int \limits_0^\infty r^2 \varepsilon \left( \phi, \chi \right)
	dr = \dfrac{4 \pi}{\tilde g^2} \hbar c m_1
	\int \limits_0^\infty x^2
	\tilde \varepsilon \left( \tilde \phi, \tilde \chi \right) dx =
	\dfrac{4 \pi}{\tilde g^2} \hbar c m_1 \tilde E
\label{mg-20}
\end{equation}
where the dimensionless quantity $\tilde \phi = \phi/m_1$, $\tilde \chi = \chi/m_1$, $x = r m_1$, $\tilde g^2 = g^2\hbar c$ are introduced; the dimension $[m_1] = $m$^{-1}$. The numerical calculation gives the following estimation $\tilde E \approx 10^{-1}$. Expression \eqref{mg-20}  allows us to estimate the value of the dimensionless coupling constant $\tilde g$ that is required for this model scalar glueball if we know the glueball mass and its characteristic size $r_0$. We will take $E_{gl} \approx 1.5\cdot 10^3$ Mev (that is an expected value of the glueball) and characteristic dimensions of glueball of the same order as the size of a proton or neutron, namely $r_0 \approx 1$  fm. Then \eqref{mg-20} gives us
\begin{equation}
	\tilde g^2 \approx 4 \pi \tilde E
	\dfrac{\hbar c m_1}{E_{gl}} \approx 1
\label{mg-30}
\end{equation}
which is in the excellent agreement with our statement that we are in the nonperturbative region.

Now we want to show that $E_{gl}$ is indeed a mass gap. In the spherical symmetric case the field equations for scalar  fields $\phi, \chi$ (that follows from the Lagrangian \eqref{mg-10}) have the form
\begin{eqnarray}
	\tilde \phi'' + \frac{2}{x} \tilde \phi' &=& \tilde \phi \left[
		\tilde \chi^2 + \lambda_2 \left( \tilde \phi^2 - \tilde m_2^2 \right)
	\right] ,
\label{mg-40}\\
	\tilde \chi'' + \frac{2}{x} \tilde \chi' &=& \tilde \chi \left[
		\tilde \phi^2 + \lambda_1 \left( \tilde \chi^2 - 1 \right)
	\right] .
\label{mg-50}
\end{eqnarray}
where the prime is the differentiation with respect to $x$. The analysis in Ref. \cite{Dzhunushaliev:2010ab} shows that if we fix the parameters values $\lambda_{1,2}$ и $m_{1,2}$ then the boundary conditions $\phi(0)$ and  $\chi(0)$ are defined in a unique way so that the solution is regular (it means that it should has a finite energy). That is, the equations \eqref{mg-40} \eqref{mg-50} 
can be considered as a nonlinear eigenvalue problem for the eigenvalues $\phi(0)$ and  $\chi(0)$ and eigenfunctions $\phi(r)$  $\chi(r)$. It means that our calculated value of the energy $E_{gl}$ is the mass gap $\Delta \approx E_{gl}$.  

\section{The lower bound of the phase transition temperature}

We now turn to the inequality \eqref{100}. After the substitution of the values of all  variables and taking $r_0 \approx 1$ fm and $\Delta \approx E_{gl}$ we obtain the following lower bound for the transition temperature
\begin{equation}
	T_c \geq \left(
		\dfrac{\Delta}{\sigma r_0^3}
	\right)^{1/4} \approx 4 \cdot 10^{10}\mathrm{K}.
\label{tp-10}
\end{equation}
At the LHC experiments is succeeded the temperature of $10^{12} \div 10^{13}$ K. Comparing this experimental value with the value calculated in \eqref{tp-10} shows that in calculating the average energy of \eqref{60} in the nonperturbative regime should either consider the energy $E_n$ with $n>2$ or the expression \eqref{70} to estimate the phase transition temperature is too rough.

\section{Conlusions}

Thus, we have qualitatively investigated how there is a  phase transition in a gluon plasma due the temperature change. At low temperatures the partition function is the sum over the discrete energy states. These states are \textcolor{blue}{\emph{correlated quantum states describing a nonperturbatively quantized SU(3) non-Abelian gauge field.}} At high temperatures the partition function is determined by the gas of interacting gluons. The phase transition for SU(3) non-Abelian gauge field means that there is a transition from the description of quantum gluon field on the language of nonperturbative correlated quantum states to the description of this quantum field on the language of perturbative interacting gluons. Such phase transition is similar to the transition from a liquid to a gas. It is crucial that in the nonperturbative case the correlated quantum states \textcolor{blue}{\emph{are not a set of quanta.}} For the calculations we have considered a glueball that is in a thermal equilibrium with the thermostat. The lower bound of the phase transition temperature has been performed by comparing the average energy glueball (which is in thermal equilibrium with the thermostat)  calculated in the perturbative and nonperturbative regimes: $T_c \gtrsim 4 \cdot 10^{10}$  K. The actual value of this temperature is above this lower bound approximately in two orders of magnitude. For a more accurate estimation of the phase transition temperature it is necessary: (a) to do a nonperturbative calculations of the next values of energy of correlated quantum states; (b) take into account the interaction between gluons.  It is shown that this lower bound with the mass gap is connected.

We have considered a scalar model of glueball in which quantum fluctuations of SU(3)  gauge field are considered in a nonperturbative way and are described by two scalar fields. The corresponding equations for these two scalar fields are considered as a nonlinear eigenvalue problem. The eigenvalues are the boundary conditions $\phi(0), \chi(0)$ for these fields and the eigenfunctions are $\phi(r), \chi(r)$. It means that the regular solutions do not exist for other values of $\phi(0), \chi(0)$. Physically, the    regular solution (solution with finite energy) describes the distribution of quantum fluctuations of the gluon field. \textcolor{blue}{\emph{The energy of such distribution of the quantum fluctuations of gluon field is a mass gap since the solutions with other boundary conditions does not exist}}. The resulting expression for the glueball energy  is compared with conventional energy value of gluevall $\approx 1.5 \cdot 10^3$ Mev. The сomparison shows that the agreement is reached in the case that the coupling constant $\tilde g$ is in a nonperturbative regime $g \approx 1$. That is in excellent agreement with our statement that we are working in the nonperturbative region.

\section*{Acknowledgements}
I am very grateful to A. Aryngazin, V. Folomeev, J. Kunz and A. Trunev for the useful discussions. This work is supported by
a grant of VolkswagenStiftung and by a grant in fundamental research in natural sciences by the Ministry of Education and Science of Kazakhstan.

\end{document}